\documentclass[letterpaper,prl,twocolumn,preprintnumbers,showpacs]{revtex4}
\usepackage{bm,graphicx}
\usepackage{amsmath}
\usepackage{amsbsy}
\usepackage{amssymb}
\usepackage{array}

\begin{document}
\title{Magnetism, spin-wave relaxation and spiral exchange in a trilayer magnetic junction}
\author{J. E. Bunder}
\affiliation{Physics Division, National Center for Theoretical Sciences, Hsinchu 300, Taiwan}
\affiliation{Department of Physics, National Tsing-Hua University, Hsinchu 300, Taiwan}
\author{Hsiu-Hau Lin}
\affiliation{Physics Division, National Center for Theoretical Sciences, Hsinchu 300, Taiwan}
\affiliation{Department of Physics, National Tsing-Hua University, Hsinchu 300, Taiwan}
\date{\today}

\begin{abstract}
We study the non-collinear exchange coupling across a trilayer magnetic junction consisting of two ferromagnets separated by a thin dilute magnetic semiconductor containing itinerant carriers with finite spin relaxation. It is remarkable that, by increasing the spin relaxation, the critical temperature is substantially enhanced and the shape of the magnetization curve becomes more mean-field like. We attribute these interesting changes to the broken time-reversal symmetry which suppresses the oscillatory Ruderman-Kittel-Kasuya-Yosida interaction. Our argument is further strengthened by the emergence of the non-collinear spiral exchange coupling across the trilayer magnetic junction with finite spin relaxation.    
\end{abstract}

\pacs{75.40.Gb, 75.50.Dd, 76.50.+g}
\maketitle

Dilute magnetic semiconductors (DMS) combine semiconducting and magnetic properties in one material~\cite{Wolf01,Zutic04}, making them popular in the field of spintronics which aims to integrate both charge and spin degrees of freedom. DMS may be fabricated by doping III-V, (such as GaAs or GaP) or II-VI (such as CdTe or ZnSe) semiconductors with a transition metal (commonly Mn). The ferromagnetism of the impurity moments is due to indirect exchange coupling mediated by the itinerant carriers in the host semiconductor~\cite{Ohno98,Macdonald05}.

A straightforward application of a DMS is to replace the sandwiching magnetic materials in a conventional trilayer magnetic junction. For instance, in conventional magnetic tunnel junctions (MTJ), which have applications in magnetic RAM, magnetic recording heads and field sensors, the central material is a semiconductor and the two magnetic materials are ferromagnets. MTJ generally show a significant giant magnetoresistive (GMR) effect. However, if the ferromagnets are replaced by DMS materials~\cite{Tanaka01,Chiba04,Wang06}, comparable or even larger GMR effects may be observed and, depending on the choice of DMS, the GMR may persist for much higher temperatures.  

\begin{figure}
\centering
\includegraphics[width=6cm]{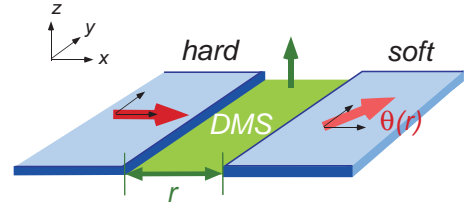}
\caption{
(Color online) Schematic diagram of the trilayer magnetic junction. The blue arrows represent the magnetization.}
\label{fig:trilayer}
\end{figure}

In this Letter, we consider a different trilayer magnetic junction where a DMS of width $r$ is sandwiched between a hard ferromagnet on the left and a soft one on the right, as shown in Fig. \ref{fig:trilayer}. We are interested in the non-collinear exchange coupling mediated by the itinerant carriers in the DMS. The spiral angle $\theta(r)$ of this trilayer junction has been studied previously, but these studies either did not consider a range of different spin relaxation rates $\gamma_h$ of the itinerant carriers~\cite{Sun04}, or they did not self-consistently derive the magnetization or chemical potential, instead using some typical values~\cite{Lin06}. Note that the self-consistent Green's function technique~\cite{Konig00,Yang01,Konig01,Bouzerar02,Sun04b,Sun06} has often been used to describe DMS, but there is little concerning the effect of the spin relaxation rate $\gamma_h$ on these calculations and other related physical properties~\cite{Bunder06}.

After including both aspects in our numerical investigations we make two important observations: (1) The critical temperature is enhanced when the finite spin relaxation $\gamma_h$ of the itinerant carriers is properly included; (2) With increasing $\gamma_h$, the usual Ruderman-Kittel-Kasuya-Yoshida (RKKY)-like interaction is suppressed and a robust spiral backbone starts to emerge. 

Our calculations start with the simple Zener model\cite{Dietl00,Konig00} for DMS,
\begin{equation}
H=H_0+J\int d^3 r\: \bm{S}(r) \cdot \bm{\sigma}(r),
\end{equation}
where $H_0$ is the kinetic energy of the carriers, $J$ is the exchange coupling, $\bm{S}(r)$ describes the impurity spin density and $\bm{\sigma}(r)=\psi^{\dag}(r)(\bm{\sigma}/2)\psi(r)$ describes the carrier spin density. For simplicity we assume a parabolic band $H_0=p^2/2m^*$. Based on experimental data for GaAs~\cite{Ohno98} we choose $J=-0.15$~eVnm$^{3}$, $m^*=m_e/2$ ($m_e$ is the mass of an electron), $n_{I}$ = 1.0 nm$^{-3}$ and $n_{h}$ = 0.1 nm$^{-3}$ for the densities of impurity spins and itinerant carriers respectively. 

In the self-consistent Green's function method, two Green's functions are defined. One, $D(p, \Omega)$, describes correlations between impurity spins and defines the spin-wave propagator while the other describes correlations between impurity and carrier spins. With the addition of Callen's formula~\cite{Sun04b,Callen63} which describes the impurity magnetization in terms of the spin-wave dispersion, one can self-consistently solve the equations of motion of the Green's functions and obtain the magnetization of the impurity and carrier spins and thus the critical temperature.

\begin{figure}
\centering
\includegraphics[width=8cm]{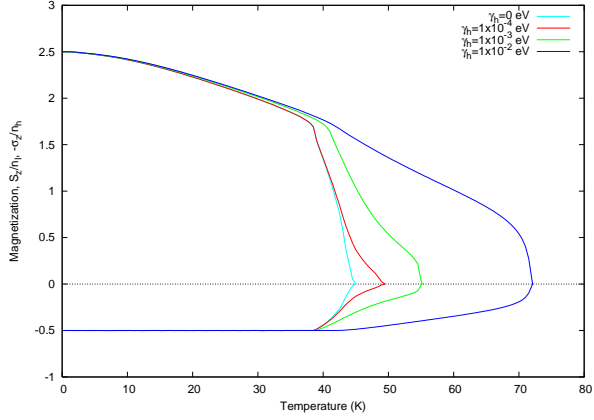}
\caption{
(Color online) The magnetization of both the impurity spins $\langle S^z\rangle$ (positive) and the inverted carrier spins $-\langle\sigma^z\rangle$ (negative) for various values of carrier relaxation rate $\gamma_h$. }
\label{fig:mag}
\end{figure}

Detailed derivations of the self-consistent Green's function are given elsewhere~\cite{Sun04b,Bunder06} so we will simply write down the relevant equations. The spin-wave propagator in the Fourier space is
\begin{eqnarray}
D(p,\Omega)&= & \frac{2\langle S_z\rangle}{\Omega-\Sigma(p,\Omega)+i \gamma_{I}},\label{eq:D2}
\end{eqnarray}
where $\gamma_I$ is the relaxation rate of the impurity spins, which we assume to be much smaller than the imaginary part of $\Sigma$ and therefore negligible. The function $\Sigma$ is the self energy of the spin waves arising from integrating out the itinerant carriers,
\begin{eqnarray}
\Sigma(p,\Omega) = -J \langle\sigma^z\rangle
+\frac{\Delta}{n_{h}}
\int \frac{d^3k}{(2\pi)^3}G(k,k+p;\Omega)
\label{eq:gf}
\end{eqnarray} 
where 
\begin{equation}
G(k, k+p;\Omega)=\frac{J n_h}{2}\frac{ f_{\uparrow}(\epsilon_k)- f_{\downarrow}(\epsilon_{k+p})}{\Omega+\epsilon_k-
\epsilon_{k+p}+\Delta+i\gamma_{h}},
\label{eq:F-D}
\end{equation}
where $f_{\alpha}(\epsilon_k)=[e^{\beta(\epsilon_k+\alpha\Delta/2-\mu)}+1]^{-1}$ is the Fermi distribution for carriers with $\alpha=\pm 1=\uparrow/\downarrow$ and the Zeeman gap $\Delta=J\langle S_z\rangle$. As $J<0$ and $\langle S^z\rangle\geq 0$ the majority band is $\alpha=\uparrow$. The spin relaxation rate of the itinerant carriers is captured by the phenomenological parameter $\gamma_h$. The equations are solved self-consistently by numerical means and the magnetizations for various $\gamma_h$ are given in Fig. \ref{fig:mag}. 

It is rather remarkable that by increasing $\gamma_h$ the critical temperature increases substantially and the general shape of the magnetization curve appears more mean field-like. This can be understood as suppression of the RKKY-like interactions which originate from quantum interferences between time-reversal symmetric patches of the Fermi surface. The presence of finite $\gamma_h$ breaks the time-reversal symmetry and thus ruins the quantum interferences. With suppressed RKKY-like interactions the frustration due to their oscillatory behavior should be absent. As a result, the critical temperature increases and the magnetization curve becomes more mean-field like.

\begin{figure}
\begin{center}
\includegraphics[width=8cm]{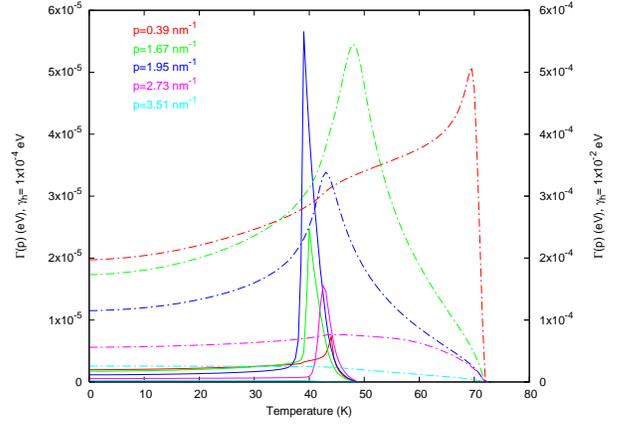}
\caption{
(Color online) The spin-wave relaxation rate $\Gamma(p)$ at different momenta and carrier relaxations rates $\gamma_h=1$x10$^{-4}$, 1x10$^{-2}$ eV. The solid (broken) lines correspond to the left (right) axis}
\label{fig:relax}
\end{center}
\end{figure}

The self-consistent Green's function approach also allows us to compute the spectral function of the spin-wave propagator $D(p,\Omega)$~\cite{Bunder06}. The spin-wave relaxation $\Gamma(p)$ is the width of the spectral function. Fig. \ref{fig:relax} gives $\Gamma(p)$ for $\gamma_h=$1$\times$10$^{-4}$, 1$\times$10$^{-2}$ eV for several momenta. As the carrier relaxation rate increases, so does the spin-wave relaxation. For $\gamma_h=$1x10$^{-4}$ the peak is sharp and tends to be in much the same position for all momenta. As $\gamma_h$ is increased the peak becomes less sharp and the lower moment peaks shift to higher temperatures.
The position and sharpness of the peaks tends to correspond to  the kinks in the magnetization curve. For smaller $\gamma_h$ we only have one sharp kink in the magnetization, but as $\gamma_h$ increases a second peak forms although neither kink is as sharp as the low $\gamma_h$ kink. 

\begin{figure}
\begin{center}
\includegraphics[width=7cm]{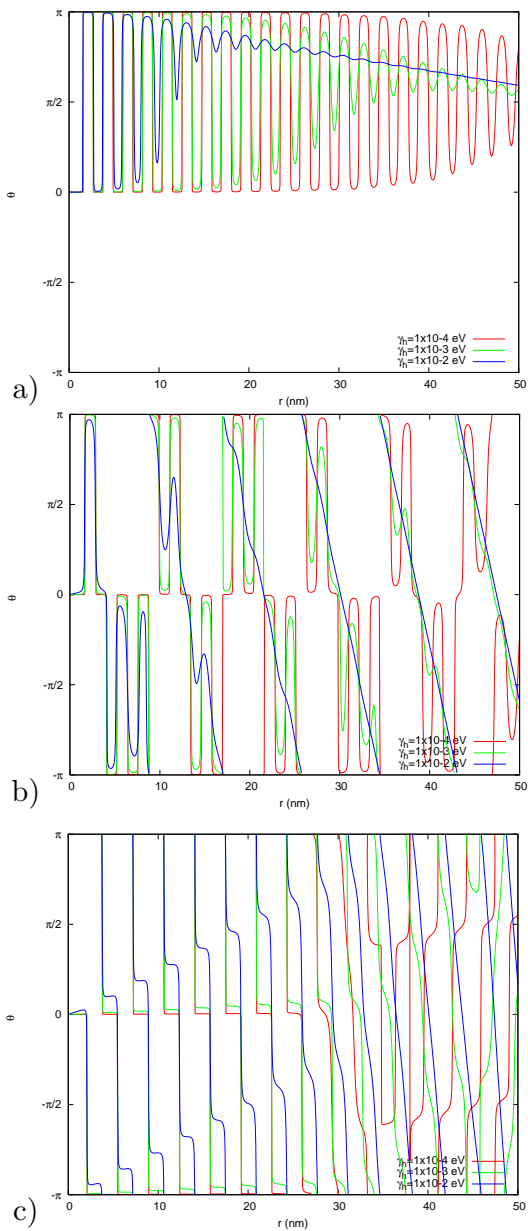}
\caption{
(Color online) The spiral angle for different $\gamma_h$ and a) $\langle S_z\rangle=$4x10$^{-2}\,n_I$, b) $\langle S_z\rangle=n_I$, c) $\langle S_z\rangle=2\,n_I$. Note that for different $\gamma_h$ the same impurity magnetization is obtained at different temperatures.} 
\label{fig:angle}
\end{center}
\end{figure}

We are now in position to compute the mediated exchange coupling across the trilayer magnetic junction shown in Fig.~\ref{fig:trilayer}. Within linear response theory~\cite{Lin06,Doniach}, it is easy to show that the effective exchange coupling is proportional to the static spin susceptibility tensor $\chi^{ij}(r)$. With the given geometry, the direction of the soft ferromagnet on the right is determined by the non-collinear angle $\theta(r) = \tan^{-1}[\chi^{yx}(r)/\chi^{xx}(r)]$. The magnetization curves obtained from the self-consistent Green's function approach can be combined with linear response theory~\cite{Lin06} to numerically calculate $\theta(r)$. Fig. \ref{fig:angle} gives the angle between ferromagnet magnetizations for different relaxation rates and impurity magnetizations.

Firstly, we consider the results for $\langle S^z\rangle <1.66\,n_I $. For small $\gamma_h$, the RKKY interaction dominates and pins the angle at $0, \pi$. As the spin relaxation $\gamma_h$ increases, the RKKY oscillation is washed out and the spiral backbone starts to emerge, particularly when $r$ is large. Furthermore, the wavelengths of the RKKY and the spiral interactions are $\lambda_{RKKY}=2\pi/(k_{F\uparrow}+k_{F\downarrow})$ and $\lambda_{s}=2\pi/(k_{F\uparrow}-k_{F\downarrow})$. As we explained in previously, the change in magnetic trends can be understood from the degree of broken time-reversal symmetry.

For $\langle S^z\rangle \geq 1.66\,n_I$, the chemical potential lies below the down-spin band and we longer have two Fermi momenta. In this regime, the results are quite different as the RKKY and spiral oscillations have the same period $2\pi/k_{F\uparrow}$. The RKKY oscillations tend to decrease in intensity with increasing $r$ and  this decay becomes slower for larger magnetizations. For large $r$, $\theta$ often displays some anomalous behavior which we do not fully comprehend yet.

In conclusion, we have shown that the spin relaxation of itinerant carriers in DMS has a significant effect on the spin-wave properties, influencing the critical temperature, the magnetization and the spectral function. We attribute these interesting changes to the broken time-reversal symmetry and find further proof in the non-collinear exchange coupling across the trilayer magnetic junction. Although we have concentrated on the magnetic properties, the transport properties of these and similar junctions should be interesting to study experimentally.

We acknowledge the grant support from the National Science Council in Taiwan through NSC 95-2112-M-007-009 and NSC 94-2112-M-007-031(HHL).


\begin{thebibliography}{99}
\bibitem{Wolf01}
S. A. Wolf, D. D. Awschalom, R. A. Buhrman, J. M. Daughton, S. von Molnar, M. L. Roukes, A. Y. Chtchelkanova and D. M. Treger, Science {\bf 294}, 1488 (2001).

\bibitem{Zutic04}
I. Zutic, J. Fabian and S. Das Sarma, Rev. Mod. Phys. {\bf 76}, 323 (2004).

\bibitem{Ohno98}
H. Ohno, Science {\bf 281}, 951 (1998).

\bibitem{Macdonald05}
A. H. MacDonald, P. Schiffer and N. Samarth, Nature Mat. {\bf 4}, 195 (2005).

\bibitem{Tanaka01}
M. Tanaka and Y. Higo, Phys. Rev. Lett. {\bf 87}, 26602 (2001).

\bibitem{Chiba04}
D. Chiba, F. Matsukura and H. Ohno, Physica E {\bf 21}, 966 (2004)

\bibitem{Wang06}
W. G. Wang, C. Ni, T. Moriyama, J. Wan, E. Nowak and J. Q. Xiao, Appl. Phys. Lett. {\bf 88}, 202501 (2006).

\bibitem{Sun04}
S.-J. Sun, S.-S. Cheng and H.-H. Lin,
Appl. Phys. Lett. {\bf 84}, 2862 (2004).

\bibitem{Lin06}
C.-H. Lin, H.-H. Lin and T.-M. Hong,
Appl. Phys. Lett. {\bf 89}, 32503 (2006).

\bibitem{Konig00}
J. Konig, H.-H. Lin and A. H. MacDonald,
Phys. Rev. Lett. {\bf 84}, 5628 (2000).

\bibitem{Yang01}
M.-F. Yang, S.-J. Sun and M.-C. Chang,
Phys. Rev. Lett. {\bf 86}, 5636 (2001).

\bibitem{Konig01}
J. Konig, H.-H. Lin and A. H. MacDonald,
Phys. Rev. Lett. {\bf 86}, 5637 (2001).

\bibitem{Bouzerar02}
G. Bouzerar and T. P. Pareek,
Phys. Rev. B {\bf 65}, 153203 (2002).

\bibitem{Sun04b}
S.-J. Sun and H.-H. Lin, Phys. Lett. A {\bf 327}, 73 (2004).

\bibitem{Sun06}
S.-J. Sun and H.-H. Lin, Eur. Phys. J. B {\bf 49}, 403 (2006).

\bibitem{Bunder06}
J. E. Bunder, S.-J. Sun and H.-H. Lin,
Appl. Phys. Lett. {\bf 89}, 72101 (2006). 

\bibitem{Dietl00}
T. Dietl, H. Ohno, F. Matsukura, J. Cibert, D. Ferrand,
Science {\bf 287}, 1019 (2000).

\bibitem{Callen63} 
Herbert B. Callen, Phys. Rev. {\bf{130}}, 890 (1963).

\bibitem{Doniach}
S. Doniach and E. H. Sondheimer, {\it Green's Functions for Solid State Physicists}, chap. 7 (World Scientific, London, 1998)

\end{thebibliography}
\end{document}